\documentclass[aps,prl,twocolumn,amsmath,amssymb,floatfix]{revtex4-2}

\usepackage{graphicx}
\usepackage{rotating}
\usepackage{color}
\usepackage{soul}
\usepackage[colorlinks,allcolors=blue]{hyperref}
\usepackage{threeparttable}
\usepackage{natbib} 
\usepackage{longtable}
\usepackage{appendix}
\usepackage{array}   
\usepackage{tabularx}
\usepackage{dcolumn}

\usepackage{booktabs}    
\usepackage{multirow}    
\usepackage{amsmath}     

\begin{document}

\title{High-Precision Measurement of D($\gamma$,\,$n$)$p$ Photodisintegration Reaction and Implications for Big-Bang Nucleosynthesis}

\author{Y. J.~Chen$^{1}$}
\author{Z. R.~Hao$^{2}$}
\author{J. J.~He$^{1}$}
\email{Corresponding author: hejianjun@fudan.edu.cn}
\author{T.~Kajino$^{3,4,5}$}
\email{Corresponding author: kajino@buaa.edu.cn}
\author{S.-I.~Ando$^{6,7}$}
\author{Y.~Luo$^{8}$}
\author{H. R.~Feng$^{3}$}
\author{L. Y.~Zhang$^{9}$}
\author{G. T.~Fan$^{2}$}
\email{Corresponding author: fangongtao@zjlab.org.cn}
\author{H. W.~Wang$^{2}$}
\author{H.~Zhang$^{1}$}
\author{Z. L.~Shen$^{9}$}
\author{L. X.~Liu$^{2}$}
\author{H. H.~Xu$^{2}$}
\author{Y.~Zhang$^{2}$}
\author{P.~Jiao$^{2}$}
\author{X. Y.~Li$^{1}$}
\author{Y. X.~Yang$^{2}$}
\author{S.~Jin$^{10}$}
\author{K. J.~Chen$^{10}$}
\author{W. Q.~Shen$^{2}$}
\author{Y. G.~Ma$^{1,11,12}$}
\email{Corresponding author: mayugang@fudan.edu.cn}

\affiliation{$^1$Key Laboratory of Nuclear Physics and Ion-beam Application (MoE), Institute of Modern Physics, Fudan University, Shanghai 200433, China}
\affiliation{$^2$Shanghai Advanced Research Institute, Chinese Academy of Sciences, Shanghai 201210, China}
\affiliation{$^3$School of Physics, Peng Huanwu Collaborative Center for Research and Education, and International Research Center for Big-Bang Cosmology and Element Genesis, Beihang University, Beijing 100191, China}
\affiliation{$^4$Graduate School of Science, The University of Tokyo, 7-3-1 Hongo, Bunkyo-ku, Tokyo 113-033, Japan}
\affiliation{$^5$Division of Science, National Astronomical Observatory of Japan, 2-21-1 Osawa, Mitaka, Tokyo 181-8588, Japan}
\affiliation{$^6$Department of Physics Education, Daegu University, Gyeongsan, Gyeongbuk 38453, Korea}
\affiliation{$^7$Department of Display and Semiconductor Engineering, Research Center for Nano-Bio Science, Sunmoon University, Asan, Chungnam 31460, Korea}
\affiliation{$^8$School of Physics, Peking University, and Kavli Institute for Astronomy and Astrophysics, Peking University, Beijing 100871, China}
\affiliation{$^9$College of Physics and Astronomy, Beijing Normal University, Beijing 100875, China}
\affiliation{$^{10}$Shanghai Institute of Applied Physics, Chinese Academy of Sciences, Shanghai 201800, China}
\affiliation{$^{11}$Shanghai Research Center for Theoretical Nuclear Physics, NSFC and Fudan University, Shanghai 200438, China}
\affiliation{$^{12}$School of Physics, East China Normal University, Shanghai 200062, China}

\date{\today}

\begin{abstract}
We report on a high-precision measurement of the D($\gamma$,\,$n$)$p$ photodisintegration reaction at the newly commissioned Shanghai Laser Electron Gamma Source (SLEGS), employing a quasi-monochromatic $\gamma$-ray beam from Laser Compton Scattering. The cross sections were determined over $E_\gamma$=2.327--7.089 MeV, achieving up to a factor of 2.2 improvement in precision near the neutron separation threshold. Combined with previous data in a global Markov chain Monte Carlo (MCMC) analysis using dibaryon effective field theory, we obtained the unprecedentedly precise $p$($n$,\,$\gamma$)D cross sections and thermonuclear rate, with a precision up to $\approx$4 times higher than previous evaluations. Implemented in a standard Big-Bang Nucleosynthesis (BBN) framework, this new rate decreases uncertainty of the key cosmological parameter of baryon density $\Omega_b h^2$ by up to $\approx$16\% relative to the LUNA result. A residual $\approx$1.2$\sigma$ tension between $\Omega_b h^2$ constrained from primordial D/H observations and CMB measurements persists, highlighting the need for improved $dd$ reaction rates and offering potential hints of new physics beyond the standard model of cosmology.

\end{abstract}

\maketitle

The hot Big-Bang theory was first proposed in 1946 by George Gamow~\cite{gam46}, and is now the most widely accepted cosmological model of the universe. According to the Big-Bang theory, the universe began with a fireball approximately 13.8 billion years ago. Following inflation and cooling, primordial Big-Bang nucleosynthesis (BBN) began when the universe was approximately 3 minutes old (when the temperature was reduced down to  approximately 1~GK), and ended less than half an hour later when nuclear reactions were quenched by the low temperature and density conditions in the expanding universe. Only the lightest nuclides were synthesized in appreciable quantities through BBN, approximately 75\% $^1$H, 25\% $^4$He, with a tiny amount of $^2$H, $^3$He and $^7$Li. These relics provide us with a unique window into the early universe~\cite{pos10,fie11,cyb16}.

In general, the primordial abundances of deuterium $^2$H (frequently written as D) and $^4$He inferred from observational data agree with the predictions of the BBN model, except for the cosmological lithium problem~\cite{fie11,cyb16}. Therefore, BBN has long been considered as one of the three historical pillars of the cosmological `Big-Bang' model. Up to now, among all primordial light elements, deuterium is the most constraining one since both its astronomical observations and BBN predictions reach about a-percent precision. In particular, deuterium is a very fragile isotope that can only be destroyed after BBN throughout stellar evolution, and its most primitive abundance is determined from the observation of cosmological clouds at high redshift, on the line of sight of distant quasars. In 2018, the precision in primordial deuterium observations was improved dramatically, reaching an accuracy of 1.19\%, {\it i.e.}, D/H=(2.527$\pm$0.030)$\times$10$^{-5}$~\cite{coo18}. In 2024, PDG recommended a newer value of (2.547$\pm$0.025)$\times$10$^{-5}$ based on the most precise observations~\cite{ParticleDataGroup:2024cfk}, reaching a precision of one percent.
In BBN, deuterium is produced through $p$($n$,\,$\gamma$)D, and subsequently destructed by three nuclear reactions of D($p$,\,$\gamma$)$^3$He, D($d$,\,$n$)$^3$He, and D($d$,\,$p$)$^3$H. These four reactions together with the neutron lifetime are the major sources of nuclear-physics uncertainty for the predicted deuterium abundance~\cite{cyb04,coc04,ser04,cyb16,pit21,shen24}. 
Therefore, it is essential to determine their cross sections at the BBN energies with a similar high precision to further constrain the predicted deuterium abundance D/H as well as the key cosmological parameter of baryon density $\Omega_b h^2$~\cite{pit18}.
This makes it necessary to reevaluate robustness of the standard BBN model in the context of precision cosmology.

For the first primordial synthesizing reaction of $p$($n$,\,$\gamma$)D (also referred to as $np$$\rightarrow$$d\gamma$~\cite{ando06}), Suzuki {\it et al.}~\cite{suzuki95} and Nagai {\it et al.}~\cite{nagai97} measured the capture cross sections at $E_n$=0.02, 0.04, 0.064 and 0.55 MeV with a precision of 5--10\% by using the pulsed neutron beams; while for its inverse process D($\gamma$,\,$n$)$p$ (also referred to as $d\gamma$$\rightarrow$$np$), 
Bishop {\it et al.}~\cite{bishop50} and Moreh {\it et al.}~\cite{more89} measured the photodisintegration cross sections at 3 energy points of $E_\gamma$=2.504, 2.618 and 2.757~MeV by using the radioactive $\gamma$-ray sources with a precision of 4--7\%; Hara {\it et al.}~\cite{hara03} measured at 7 energy points over $E_\gamma$=2.33--4.58~MeV with a precision of 6--10\% by using the Laser Compton Scattering (LCS) $\gamma$-ray beams; other miscellaneous experimental information can be referred to Refs.~\cite{bire85,gra92,sch00,tor03,ser04,BOOK}.
It should be noted that directly applying these experimental data to the BBN predictions would increase the uncertainties. 
The effective field theory (dEFT) calculations suggest that theory errors can be sufficiently smaller than the experimental ones. It can provide a very useful discriminant for theories and their perturbative schemes. The compilations of BBN reactions adopt these theory-based cross sections because they can provide more robust and accurate predictions than experiments alone. However, the uncertainties from the recent theoretical estimations of the cross section for $np$$\rightarrow$$d\gamma$ at BBN energies are considerably different from each other, {\it e.g.}, $\approx$4\% in~\cite{chen99}, 2--3\% in~\cite{joh01}, $\approx$1\% in~\cite{rup00}, and $\leq$1\% in~\cite{ando06}. These differences could lead to different uncertainties in the BBN predictions. Therefore, more precise experimental data in the BBN energy region are strongly required to characterize the nuclear reaction model. In addition, predictions of the $R$-matrix theory at $E_\mathrm{c.m.}$=0.1 and 1 MeV were found to deviate from others by $\approx$4.6\%~\cite{ando06}, which also require further verification.

A new type of LCS $\gamma$-ray source, named Shanghai Laser Electron Gamma Source (SLEGS)~\cite{wang22,xu22,liu24}, has been recently commissioned in China. SLEGS is one of the beamlines of Shanghai Synchrotron Radiation Facility (SSRF)~\cite{he14}. It uses a CO$_2$ laser~\cite{xu2025} to interact with the 3.5~GeV electrons from the SSRF storage ring. This interaction can be implemented via both the slant-scattering and back-scattering modes, and generates quasi-monochromatic $\gamma$-ray beams in an energy range of 0.4--21.7~MeV, with a flux of 10$^5$--10$^7$~photons/s via a collimation system~\cite{hao21}. SLEGS is the first LCS facility to produce the high-flux $\gamma$-ray beam in a laser Compton slant-scattering mode ({\it i.e.}, LCSS mode), which can provide more convenient energy scanning capability.

In this Letter, we report the results of a high-precision experiment for the D($\gamma$,\,$n$)$p$ photodisintegration reaction at SLEGS. The photoneutron cross sections were measured at 22 energy points near the neutron threshold. Our new cross sections are up to a factor of 2.2 more precise than the previous ones~\cite{hara03}. The cross sections of $p$($n$,\,$\gamma$)D have been evaluated by dEFT with a Markov chain Monte Carlo (MCMC) analysis, together with our new data and all other relevant experimental data. With the more precise cross sections evaluated, an unprecedentedly precise thermonuclear $p$($n$,\,$\gamma$)D rate is thus obtained, approximately 4 times more precise than the previous ones~\cite{ando06,ser04} in the BBN temperature regime. The impact of the present high-precision $p$($n$,\,$\gamma$)D rate has been investigated with a standard BBN model, and astrophysical implications are discussed on the cosmological parameter of baryon density $\Omega_bh^2$.

The experiment was carried out at the SLEGS facility, and the experimental setup is similar to that described in~\cite{hao25}. The quasi-monochromatic $\gamma$-ray beam was generated via the LCSS mode, achieved by the interaction of the electron beam from the SSRF and the CO$_2$ laser. The laser was operated at a power of 5~W, with a frequency of 1~kHz and pulse width of 50~$\mu$s~\cite{xu2025}. The $\gamma$-ray beam was collimated by a $\phi$5~mm coarse collimator and a $\phi$2~mm fine one (C5T2)~\cite{hao21}. The collimated $\gamma$-rays irradiated a high purity D$_2$O (99.9\%) water target that was sealed in a pure aluminium container ($\phi$=10~mm, $L$=100~mm, with an equivalent deuterium areal density of $N_T$=6.65$\times$10$^{23}$/cm$^2$) at 22 different laser--electron collision angles. The weighted average $\gamma$-ray energies ranged from $E_\gamma^\mathrm{WA}$=2.327--7.089~MeV (in quasi-uniform energy steps) with a flux of approximately 10$^{5}$ photons/s. The residual $\gamma$-ray beam was attenuated by an external copper attenuator with a thickness ranging from 160~mm to 175~mm, which was monitored by a LaBr$_3$ detector ($\phi$3$\times$4~inch) in real time. Similarly to Refs.~\cite{liu24,hao25}, the incident $\gamma$-beam energy spectra on the D$_2$O target were reconstructed by those of the LaBr$_3$ detector accordingly (see Supplemental Material~\cite{supp}). Here, the target was aligned in the geometric center of a 4$\pi$ flat-efficiency $^3$He neutron detector (FED)~\cite{hao20,hao25}, where the photoneutrons were moderated with the surrounding polyethylene material and subsequently captured by the $^3$He counters.

Similarly to the previous analysis method~\cite{ren18,hao25}, the measured photoneutron cross section for an incident $\gamma$-ray beam with maximum energy of $E_\mathrm{max}$ can be expressed by:
\begin{equation}
		\sigma_\mathrm{exp}^{{E_\mathrm{max}}}(E_\gamma^\mathrm{WA}) = \int^{E_\mathrm{max}}_{S_n} D^{E_\mathrm{max}}(E_\gamma)\sigma(E_\gamma) dE_\gamma = \frac{N_n}{N_T N_\gamma \xi \epsilon_n g}.
\label{eq1:fold}
\end{equation}

\noindent
Where $D^{E_\mathrm{max}}$ is the normalized, $\int^{E_\mathrm{max}}_{S_n} D^{E_\mathrm{max}} dE_\gamma$=1, energy distribution of the $\gamma$-ray beam obtained using the method described in Ref.~\cite{liu24}. Here, $\sigma(E_\gamma)$ represents the true monochromatic photoneutron cross section as a function of energy $E_\gamma$. $N_n$ is the number of photoneutrons detected by the FED, and $N_\gamma$ is the number of $\gamma$ photons incident on the target. $\epsilon_n$ is the average detector efficiency determined using the Ring-Ratio (\textit{RR}) technique~\cite{hao20,hao25}. $N_T$ is the number of target nuclei per unit area. $\xi$=$(1 - e^{-\mu t}) / (\mu t)$ is a correction factor for a thick target, where $\mu$ is the mass attenuation coefficient, and $t$ is the target thickness. $g$ is the fraction of the $\gamma$-ray flux above the neutron separation energy of $S_n$=2.2246 MeV~\cite{ame20}. Here, we define a weighted average $\gamma$-ray beam energy ($E_\gamma^\mathrm{WA}$) as follows:
\begin{equation}
E_\gamma^\mathrm{WA}=\frac{\int_{S_n}^{E_\mathrm{max}}E_\gamma D^{E_\mathrm{max}}(E_\gamma)dE_\gamma}{\int_{S_n}^{E_\mathrm{max}}D^{E_\mathrm{max}}(E_\gamma)dE_\gamma}.
\label{eq2:Eave}
\end{equation}

In fact, the quasi-monochromatic cross section $\sigma_\mathrm{exp}^{E_\mathrm{max}}$ introduced above is a folded or convoluted cross section over the whole energy range of the incident $\gamma$-ray beam, which is defined here as $\sigma^\mathrm{f}$ (also in Ref.~\cite{ren18}). The uncertainties in $\sigma^\mathrm{f}$ consist mainly of statistical, systematic and methodological ones. Thereinto, statistical uncertainties $\Delta\sigma_\mathrm{stat}$ are within 0.5–1.6\% for most data points, with only two data points closest to the neutron threshold having larger statistical uncertainties of 3.6\% and 10\%, respectively. The systematic uncertainty $\Delta\sigma_\mathrm{sys}$ is mainly attributed to the following two factors~\cite{hao25_1,hao25}: 1) the FED efficiency ($\approx$3.0\%); 2) the reconstructed incident $\gamma$-ray energy spectrum owing to the copper attenuator and the detection efficiency of LaBr$_3$ ($\approx$2.0\%). The total systematic uncertainty is estimated to be approximately 3.6\%. Furthermore, the methodological uncertainty $\Delta\sigma_\mathrm{meth}$~\cite{haoPRL_sub} in the neutron extraction and $\gamma$ spectral unfolding procedure is estimated to be 1.6--1.9\%. For the total uncertainty, we have assumed that all the uncertainties estimated above are independent, and thus they were added quadratically. The folded photoneutron data are listed in the Supplemental Material~\cite{supp}.

\begin{figure}[h]
    \centering
    \includegraphics[width=8cm]{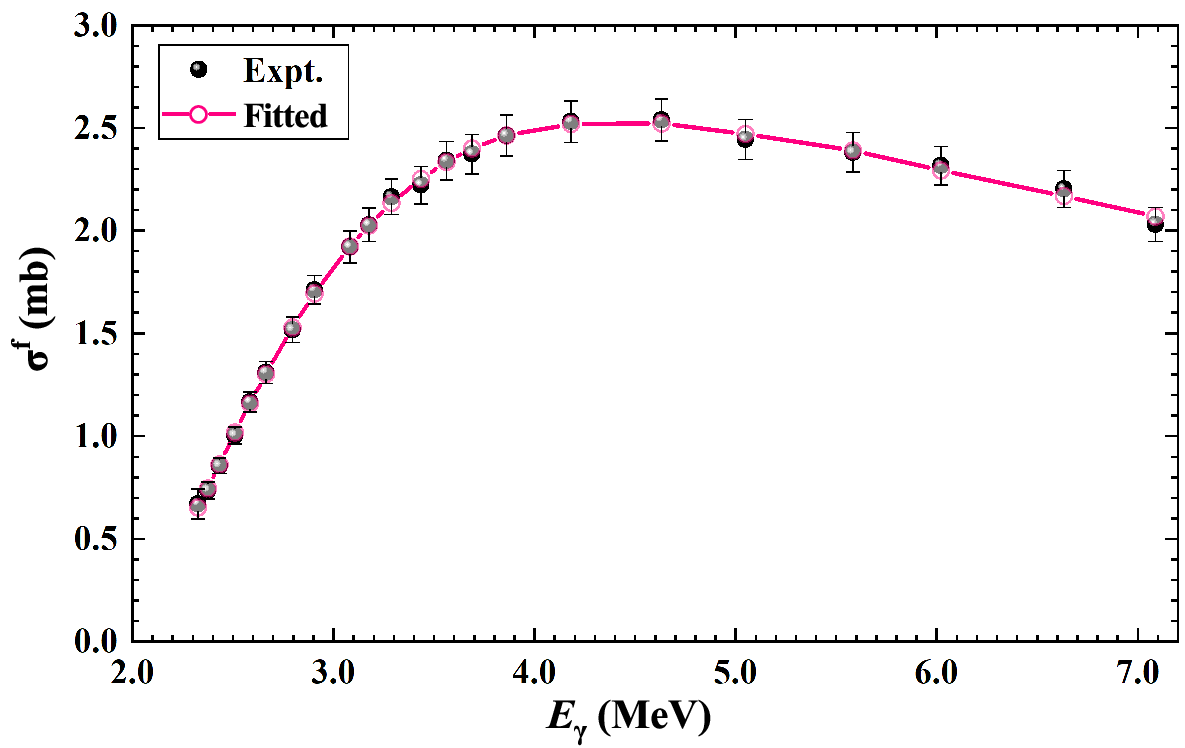}
    \vspace{-3mm}
    \caption{The 4$\pi\cdot A_0$ fit to the folded cross sections (median values) of D($\gamma$,\,$n$)p measured with SLEGS. The experimental data and and fitted ones are indicated by the solid points and circles, respectively. The solid curve connecting the fitted data points is just for guiding the eyes.}
    \label{fig:A0Fit}
\end{figure}

\begin{figure}[h!]
    \centering
    \includegraphics[width=8.4cm]{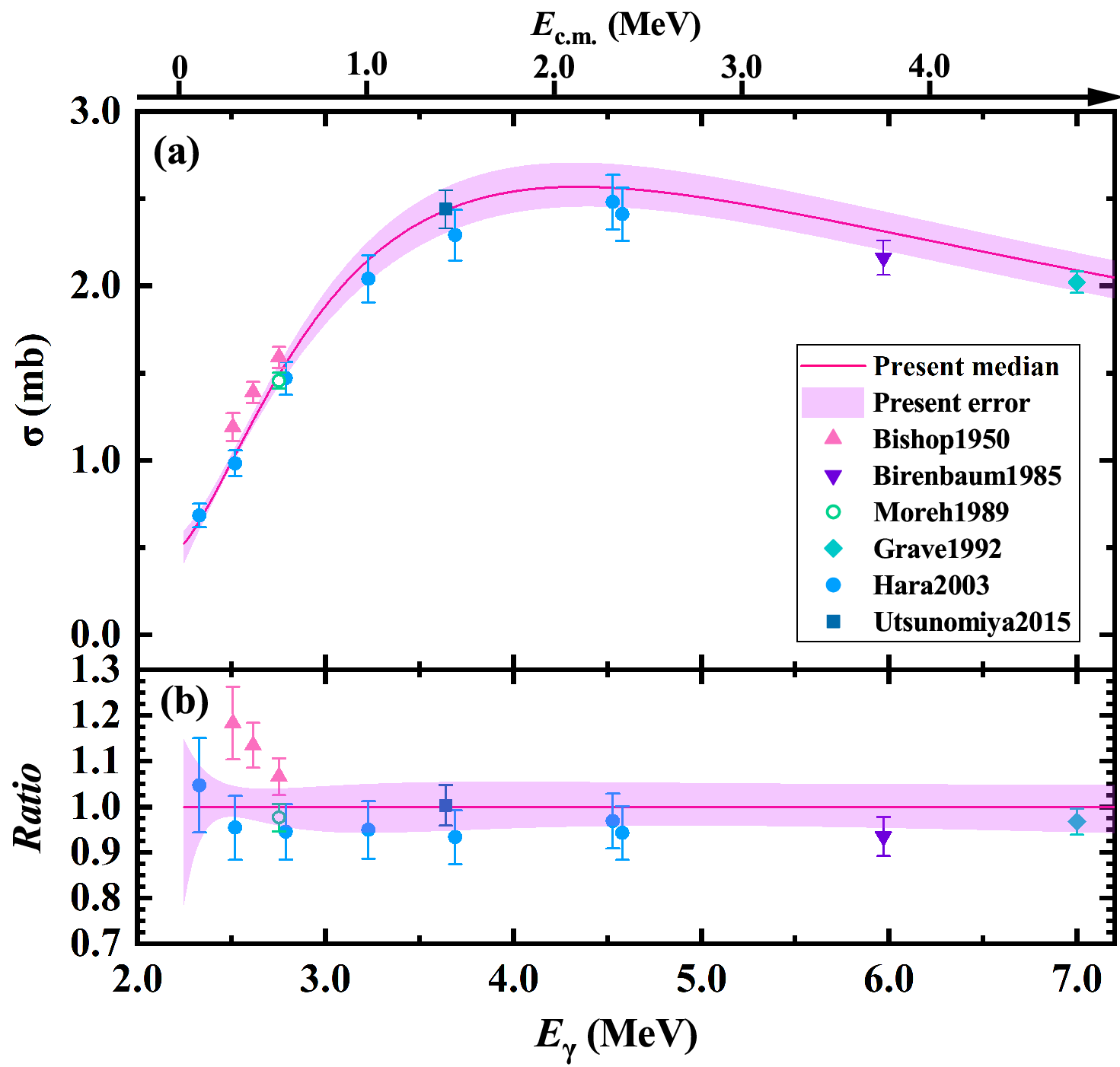}
    \vspace{-2mm}
    \caption{(a) Photoneutron cross sections of D($\gamma$,\,$n$)$p$ reaction. The present SLEGS results are shown as a red line, with the associated uncertainties indicated by a colored band. The previous data~\cite{bishop50,bire85,more89,gra92,hara03,uts15} are shown for comparison; (b) Ratios of cross sections between the present results (as the reference) and previous ones. See more details in the text.}
    \label{fig:mono-cross-section}
\end{figure}

The cross sections available for the D($\gamma$,\,$n$)$p$ photodisintegration can be well described by a phenomenological expression~\cite{BOOK}, {\it i.e.}, 
\begin{equation}\label{eq3:sigma}
    \sigma(E_\gamma) = 4\pi \times A_0(E_{\gamma})
\end{equation}
with a parameterized Legendre polynomial coefficient
\begin{equation}\label{eq4:A0}
    A_{0}(E_{\gamma}){=}c_{1}e^{c_{2}E_{\gamma}}+c_{3}e^{c_{4}E_{\gamma}}+\frac{c_{5}+c_{6}E_{\gamma}}{1+c_{8}(E_{\gamma}-c_{7})^{2}}.
\end{equation}
\noindent
We have convoluted this parameterized cross section with the $\gamma$-beam energy spectrum and performed least-squares fits to the 22 experimental folded data points using Eq.~\ref{eq1:fold}. The folded cross sections ($\sigma^\mathrm{f}$) and the associated uncertainties (i.e., the lower and upper limits) can be well fitted by Eq.~\ref{eq3:sigma}. As an example, Fig.~\ref{fig:A0Fit} illustrates the corresponding fit to the median values of folded cross sections. All the corresponding parameters $c_i$ are listed in the Supplemental Material~\cite{supp}, which can be used to directly calculate the (unfolded) monochromatic photoneutron cross sections and their uncertainties in the energy region of $E_\gamma$=2.25--7.20~MeV with Eqs.~3\&4 (see the colored band in Fig.~\ref{fig:mono-cross-section}).

Figure~\ref{fig:mono-cross-section} compares our (unfolded) monochromatic photoneutron cross sections with the previous ones. Our new values are a factor of 1.2--2.2 more precise than Hara {\it et al.}'s data~\cite{hara03}, both with a similar experimental method. It should be noted that Hara {\it et al.}'s data are $\approx$5.2\% systematically lower than the present ones (except for their lowest energy point), possibly because they are a kind of quasi-monochromatic or folded cross sections. Therefore, our precise measurements over a wide energy region (with better consistency) can act as a better reference for evaluating the $p$($n$,\,$\gamma$)D cross sections.

\begin{table}[h!]
\footnotesize
\centering
\caption{Cross sections ($\sigma_\mathrm{tot}$) evaluated for the $p$($n$,\,$\gamma$)D reaction at some BBN relevant energies together with the $M$1 and $E$1 contributions ($\sigma_{M1}$, $\sigma_{E1}$). The previous evaluations~\cite{ando06} are listed comparison. Here, the uncertainties are listed in the parentheses.}
\label{tab:cross_sections}
\begin{ruledtabular}
\begin{tabular}{ccccc}
$E_{\mathrm{c.m.}}$ & \multicolumn{3}{c}{Present} & \multicolumn{1}{c}{Ref.~\cite{ando06}} \\
\cmidrule{2-4}
(MeV) & $\sigma_{M1}$(mb) & $\sigma_{E1}$(mb) & $\sigma_\mathrm{tot}$(mb) & $\sigma_\mathrm{tot}$(mb) \\
\midrule
1.265$\times$10$^{-8}$ & 333.5(4)    & 5.09(1)$\times$10$^{-6}$ & 333.5(4)     & 333.8(15) \\ 
0.0005     & 1.665(2)    & 0.001012(2) & 1.666(2)     & 1.667(8)      \\ 
0.001     & 1.169(1)    & 0.001431(3) & 1.171(1)  & 1.171(5)   \\ 
0.005     & 0.4945(6)   & 0.00319(1) & 0.4977(6)    & 0.4979(23)   \\ 
0.01     & 0.3276(4)   & 0.00451(1) & 0.3321(4)    & 0.3322(15)   \\ 
0.05     & 0.0981(1)   & 0.00990(2) & 0.1080(1)    & 0.1079(5)    \\ 
0.1                   & 0.04969(7)  & 0.01370(3) & 0.06339(7)   & 0.0634(3) \\ 
0.2                   & 0.02300(4)  & 0.01858(4) & 0.04158(5)   & - \\ 
0.3                   & 0.01428(2)  & 0.02185(5) & 0.03614(5)   & - \\ 
0.4                   & 0.01015(2)  & 0.02427(5) & 0.03442(6)   & - \\ 
0.5                   & 0.00780(1)   & 0.02614(6) & 0.03394(6)   & 0.0341(2) \\ 
0.6                   & 0.00630(1)  & 0.02762(6) & 0.03392(6)   & - \\ 
0.7                   & 0.00528(1)  & 0.02881(6) & 0.03409(7)   & - \\ 
0.8                   & 0.00454(1)  & 0.02978(7) & 0.03432(7)   & - \\ 
0.9                   & 0.00398(1)  & 0.03058(7) & 0.03456(7)   & - \\ 
1.0                   & 0.00355(1)  & 0.03123(7)  & 0.03478(7)    & 0.0349(3)   \\ 
\end{tabular}
\end{ruledtabular}
\end{table}

\begin{figure}[htbp]
    \centering
    \includegraphics[width=8.4cm]{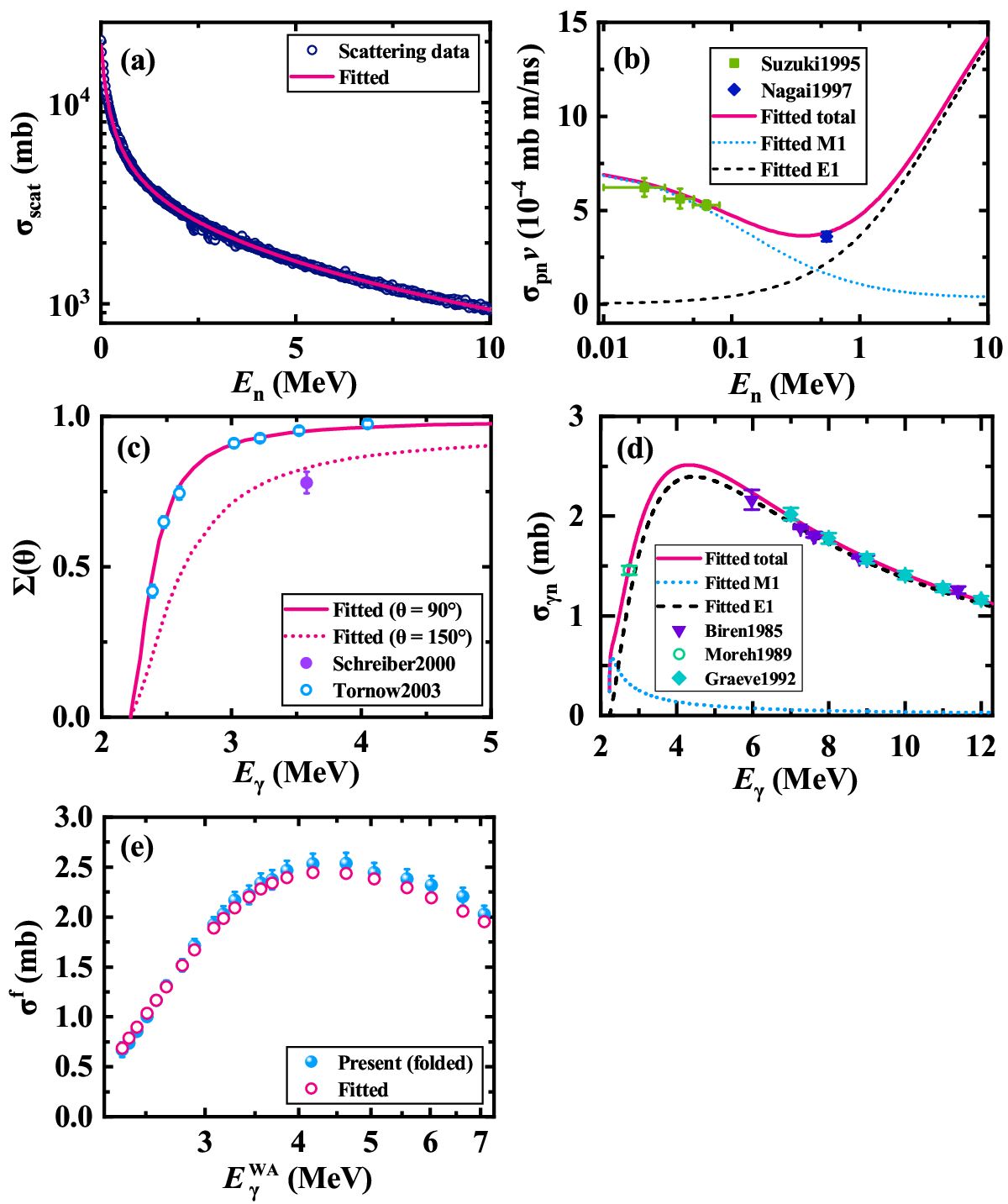}
    \vspace{-2mm}
    \caption{MCMC fits to experimental data available in following four channels: (a) $np$ scattering cross section~\cite{nn-online}; (b) $np$ capture cross section multiplied by neutron speed~\cite{suzuki95,nagai97}; (c) photon analyzing power $\Sigma$($\theta$) for $d\vec{\gamma}$$\rightarrow$$np$ process~\cite{sch00,tor03}; (d) cross section of D($\gamma$,\,$n$)$p$~\cite{bire85,more89,gra92}; (e) folded data of D($\gamma$,\,$n$)$p$ obtained at SLEGS. See more details in Ref.~\cite{ando06}.}
    \label{fig:ando_fit}
\end{figure}

Similar to the method used in Ref.~\cite{ando06}, a global fitting within the framework of the dEFT has been performed based on the Markov Chain Monte Carlo (MCMC) analysis. All the relevant experimental data, {\it i.e.}, the $np$ scattering cross sections in the neutron energies of $E_n$$\leq$10~MeV~\cite{nn-online}, the $np$$\rightarrow$$d\gamma$ capture cross sections~\cite{suzuki95,nagai97}, the photon analyzing power for the $d\vec{\gamma}$$\rightarrow$$np$ process~\cite{sch00,tor03}, the photoneutron cross sections~\cite{bire85,more89,gra92}, as well as our folded cross sections of D($\gamma$,\,$n$)$p$ obtained at SLEGS, have been included in the present evaluation. The results of the global MCMC fitting are shown in Fig.~\ref{fig:ando_fit}. Compared to the previous evaluation~\cite{ando06}, the present SLEGS data together with the recently measured $np$ scattering data~\cite{DA13} and the photoneutron data measured at the higher energies of $E_\gamma$$\approx$6--12~MeV~\cite{bire85,gra92}, have been newly included here, without including Hara {\it et al.}'s~\cite{hara03} data. Similar to Ref.~\cite{ando06}, since Bishop {\it et al.}'s~\cite{bishop50} data were found to have a problem~\cite{nagai97}, they are also not included in the present evaluation. The presently evaluated $np$ capture cross sections together with the $M$1 and $E$1 contributions are listed in Table~\ref{tab:cross_sections}. It shows that our evaluated cross sections are almost identical to the previous ones~\cite{ando06} (at most 0.5\% difference), while the present uncertainties are significantly reduced to 0.10--0.21\% in the energy region of 0.01--1.0 MeV, {\it i.e.}, about 3.3--4.3 times more precise. It is found again that Hara {\it et al.}'s data are $\approx$4.2\% systematically lower for the reason discussed above. In addition, significant deviations can still be observed for the $R$-matrix results~\cite{joh01} at 0.05, 0.1 and 1 MeV points (up to $\approx$5.0\% at 1 MeV). The evaluation confirms that this $R$-matrix calculation still needs certain modifications. Similar to previous evaluations~\cite{ser04,ando06}, the intrinsic error of EFT theory was also not involved here. The precise characterization of such error is still required, which is beyond the scope of this Letter.

A new $np$ capture rate in the temperature range of 0.01--10 GK has been calculated using the presently evaluated $p$($n$,\,$\gamma$)D cross sections, and its upper and lower limits estimated using the corresponding uncertainties of the cross sections. The uncertainty of the present rate is reduced significantly down to $\approx$0.12\% in the temperature regime of BBN interest (i.e., 0.1--1 GK). The present rate (median value) deviates only slightly (at most 0.23\%) from the previous Ando {\it et al.}'s evaluation~\cite{ando06}, however, the corresponding uncertainty is reduced by a factor of $\approx$4; compared to the Serpico {\it et al.}'s rate~\cite{ser04}, our rate is only slightly (at most 1\%) smaller, but a factor of 1.9--3.5 times more precise (see Supplemental Material~\cite{supp}).
Thus, we recommend here an unprecedentedly precise rate for the $p$($n$,\,$\gamma$)D reaction, which is of great importance for precision cosmological studies.

The impact of our high-precision $p$($n$,\,$\gamma$)D rate has been investigated with a standard $\Lambda$CDM BBN model~\cite{Kawano1992,Smith:1992yy}, with baryon density $\Omega_bh^2$ as a single free parameter, by performing a Bayesian analysis. Thereinto, the neutron lifetime is taken as $\tau$=878.4$\pm$0.5~s~\cite{ParticleDataGroup:2024cfk}, and other 3 essential rates on determining D/H, namely, D$(d,p)^3\rm H$, D$(d,n)^3\rm He$ and D$(p,\gamma)^3$He, are taken from Ref.~\cite{Pisanti:2020efz}. Using a deuterium abundance observed by Cooke {\it et al.}~\cite{coo18}, we place a constraint of $\Omega_bh^2$=0.02231$\pm$0.00032 in the standard $\Lambda$CDM model with effective neutrino species $N_{\rm eff}$=3.045~\cite{man05,sal16}, {\it i.e.}, improving the uncertainty by $\approx$11\% compared to the previous LUNA value of 0.02233$\pm$0.00036~\cite{LUNA20} (with Serpico {\it et al.}'s rate~\cite{ser04}). However, a tighter constraint of $\Omega_bh^2$=0.02220$\pm$0.00031 can be achieved based on the newest PDG recommended value~\cite{ParticleDataGroup:2024cfk}. This will improve the uncertainty of $\Omega_bh^2$ by $\approx$16\%.
Furthermore, our rate can reduce the uncertainty of $\Omega_bh^2$ by $\approx$2.5\% compared to Ando {\it et al.}'s rate~\cite{ando06}.

\begin{figure}[t]
	\begin{center}
		\includegraphics[width=7.8cm]{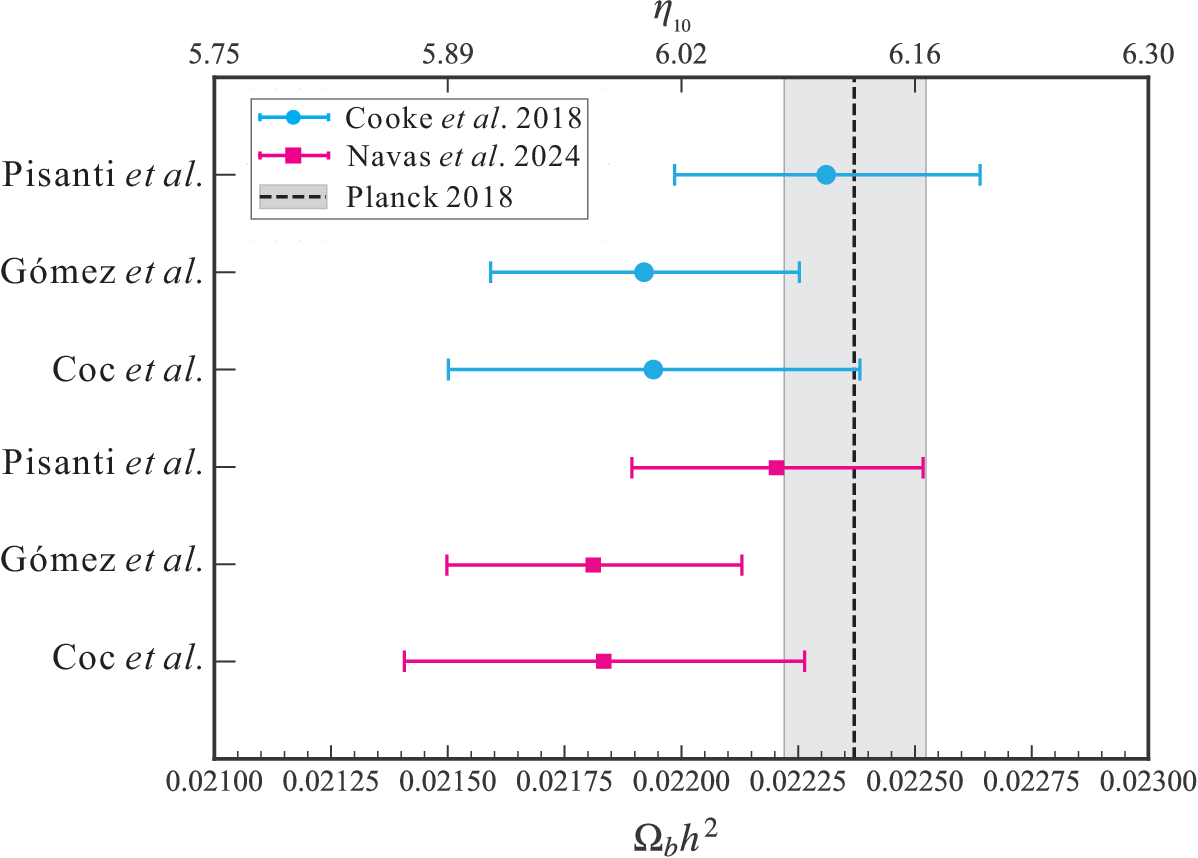}
		\caption{\label{fig5}BBN constraints on $\Omega_bh^2$ calculated for three different sets of $dd$ rates~\cite{Pisanti:2020efz,gom17,coc15}, with 1$\sigma$ error bar indicated. The constraints using observed D/H values of Cooke {\it et al.}~\cite{coo18} and of Navas {\it et al.}~\cite{ParticleDataGroup:2024cfk} are shown for comparison. The Planck CMB observation~\cite{Planck:2018vyg} is indicated by the gray band.}
	\end{center}
\end{figure}

It is known that the cross section of D$(p,\gamma)^3$He measured by the recent LUNA experiment~\cite{LUNA20} has already reached an unprecedented precision of better than 3\%. Thus, only D$(d,p)^{3}$H and D$(d,n)^{3}$He reactions (hereafter short for $ddp$ and $ddn$, respectively) now constitute the dominant nuclear physics uncertainties in predicting D/H and $\Omega_bh^2$, because their cross section data still have large uncertainties. To evaluate the astrophysical $S$-factors of $ddp$ and $ddn$ reactions for calculating their rates, Pisanti {\it et al.}~\cite{Pisanti:2007hk,Pisanti:2020efz} employed polynomial fits to multiple experimental datasets, while Coc {\it et al.}~\cite{coc15} and G\'omez {\it et al.}~\cite{gom17} utilized \textit{ab initio} calculations as theoretical constraints for experimental data selection. The difference between the rates of Coc {\it et al.} and G\'omez {\it et al.} is quite small (less than $\approx$0.4\%), while those of Pisanti {\it et al.} deviate from other two rates up to factors of 6.3\% and 3.3\% for $ddp$ and $ddn$ reactions in the BBN temperatures, respectively. Thereinto, errors in the rates of Pisanti {\it et al.}, G\'omez {\it et al.} and Coc {\it et al.} were assumed to be 1.0\%, 1.1\% and 2.0\%, respectively. Figure~\ref{fig5} shows the constraints on $\Omega_b h^2$ calculated using three different $dd$ rates together with Cooke {\it et al.}'s observed value~\cite{coo18}. It shows that the $\Omega_bh^2$ values constrained with both G\'omez {\it et al}. and Coc {\it et al}.'s $dd$ rates only roughly match with the Planck CMB observations~\cite{Planck:2018vyg} ({\it i.e.}, $\Omega_bh^2$=0.02237$\pm$0.00015, indicated by a gray band). However, if the newest D/H  value~\cite{ParticleDataGroup:2024cfk} is adopted, then a remarkable $\approx$1.2$\sigma$ tension in the $\Omega_bh^2$ constraints appears between the BBN prediction and the CMB observations once using the G\'omez {\it et al.}'s $dd$ rates. It shows that not only the $dd$ rates (median value and associated error), but also the observed deuterium abundance can affect this tension.

\vspace{5mm}
In summary, we have studied the key BBN reaction $p$($n$,\,$\gamma$)D via its time-reversal process D($\gamma$,\,$n$)$p$ using a quasi-monochromatic $\gamma$-ray source. The cross sections obtained are much more precise than the previous ones, allowing the evaluation of the $p$($n$,\,$\gamma$)D cross sections and reaction rate with unprecedented high precision. Within a $\Lambda$CDM BBN framework, this improvement reduces the uncertainty of the cosmological parameter $\Omega_b h^2$ by up to $\approx$16\% compared to the previous result. The dominant nuclear physics uncertainties in the predictions of D/H and $\Omega_bh^2$ now arise from $dd$ reactions, where different $dd$ rate choices can lead to a $\approx$1.2$\sigma$ tension between $\Omega_b h^2$ constraints from primordial D/H observations and CMB measurements. Resolving such tension or discrepancy requires future high-precision measurements and theoretical refinements of $dd$ reactions, which may reconcile current datasets or reveal new physics. Furthermore, this work demonstrates the capability of SLEGS to deliver precise nuclear data of astrophysical importance.

\vspace{10mm}
\begin{center}
\textbf{Acknowledgments}
\end{center}
We thank the staff of the Shanghai Synchrotron Radiation Facility (BL03SSID, https://cstr.cn/31124.02.SSRF.BL03SSID) for the assistance on this measurement. J.~J.~H. acknowledges Hiroaki Utsunomiya's contribution to data reduction. This work was financially supported by the National Key R\&D Program of China (Nos. 2022YFA1602401) and the National Natural Science Foundation of China (Nos. 12322509, 12547102, 12275338, 12335009, 12435010). Y.~L. acknowledges the support of the Boya fellowship of Peking University and the China Postdoctoral Science Foundation (No. 2025T180924). S.-I.~A. acknowledges the support of National Research Foundation of Korea (NRF) grant (Nos. RS-2025-16065411, 2022R1F1A1070060 and 2023R1A2C1003177).


\begin{thebibliography} {99}
\bibitem{gam46} 
G. Gamow, \href{https://link.aps.org/doi/10.1103/PhysRev.70.572.2}{Phys. Rev. {\bf 70}, 572 (1946)}.
\bibitem{pos10}
M. Pospelov and J. Pradler, \href{https://doi.org/10.1146/annurev.nucl.012809.104521}{Annu. Rev. Nucl. Part. Sci. {\bf 60}, 539 (2010)}.
\bibitem{fie11}
B. D. Fields, \href{https://doi.org/10.1146/annurev-nucl-102010-130445}{Annu. Rev. Nucl. Part. Sci. {\bf 61}, 47 (2011)}.
\bibitem{cyb16}
R. H. Cyburt \textit{et al}., \href{https://link.aps.org/doi/10.1103/RevModPhys.88.015004}{Rev. Mod. Phys. {\bf 88}, 015004 (2016)}.
\bibitem{coo18} 
R. J. Cooke, M. Pettini, and C. C. Steidel, \href{https://dx.doi.org/10.3847/1538-4357/aaab53}{\apj ~{\bf 855}, 102 (2018)}.
\bibitem{ParticleDataGroup:2024cfk}
S.~Navas \textit{et al.} (Particle Data Group), \href{https://doi.org/10.1103/PhysRevD.110.030001}{Phys. Rev. D {\bf 110}, 030001 (2024)}.
\bibitem{cyb04}
R. H. Cyburt, \href{https://link.aps.org/doi/10.1103/PhysRevD.70.023505}{\prd ~{\bf 70}, 023505 (2004)}.
\bibitem{coc04}
A. Coc, E. Vangioni-Flam, P. Descouvemont, A. Adahchour, and C. Angulo, \href{https://iopscience.iop.org/article/10.1086/380121}{\apj ~{\bf 600}, 544 (2004)}.
\bibitem{ser04}
P. D. Serpico \textit{et al.}, \href{https://dx.doi.org/10.1088/1475-7516/2004/12/010}{J. Cosmol. Astropart. Phys. {\bf 2004}, 010 (2004)}.
\bibitem{shen24}
Z. L. Shen and J. J. He, \href{https://doi.org/10.1007/s41365-024-01423-3}{Nucl. Sci. Tech. {\bf 35}, 63 (2024)}.
\bibitem{pit21}
C. Pitrou \textit{et al.}, \href{https://doi.org/10.1093/mnras/stab135}{Mon. Not. R. Astron. Soc. {\bf 502}, 2474 (2021)}.
\bibitem{pit18}
C. Pitrou \textit{et al.}, \href{https://www.sciencedirect.com/science/article/pii/S0370157318301054}{Phys. Rep. {\bf 754}, 1 (2018)}.
\bibitem{ando06}
S. Ando \textit{et al.}, \href{https://link.aps.org/doi/10.1103/PhysRevC.74.025809}{Phys. Rev. C {\bf 74} 025809 (2006)}.
\bibitem{suzuki95}
T. Suzuki \textit{et al.}, \href{https://ui.adsabs.harvard.edu/abs/1995ApJ...439L..59S}{\apj ~{\bf 439}, L59 (1995)}.
\bibitem{nagai97}
Y. Nagai \textit{et al.}, \href{https://link.aps.org/doi/10.1103/PhysRevC.56.3173}{Phys. Rev. C {\bf 56}, 3173 (1997)}.
\bibitem{bishop50}
G. R. Bishop \textit{et al.}, \href{https://doi.org/10.1103/PhysRev.80.211}{Phys. Rev. {\bf 80}, 211 (1950)}.
\bibitem{more89}
R. Moreh \textit{et al.}, \href{https://link.aps.org/doi/10.1103/PhysRevC.39.1247}{Phys. Rev. C {\bf 39}, 1247 (1989)}.
\bibitem{hara03}
K. Hara \textit{et al.}, \href{https://link.aps.org/doi/10.1103/PhysRevD.68.072001}{Phys. Rev. D {\bf 68}, 072001 (2003)}.
\bibitem{gra92}
A. De Graeve \textit{et al.}, \href{https://doi.org/10.1103/PhysRevC.45.860}{Phys. Rev. C {\bf 45}, 860 (1992)}.
\bibitem{sch00}
E. C. Schreiber \textit{et al.}, \href{https://doi.org/10.1103/PhysRevC.61.061604}{Phys. Rev. C {\bf 61}, 061604(R) (2000)}.
\bibitem{tor03}
W. Tornow \textit{et al.}, \href{https://doi.org/10.1016/j.physletb.2003.08.078}{Phys. Lett. B {\bf 574}, 8 (2003)}.
\bibitem{bire85}
Y. Birenbaum \textit{et al.}, \href{https://link.aps.org/doi/10.1103/PhysRevC.32.1825}{Phys. Rev. C {\bf 32}, 1825 (1985)}.
\bibitem{BOOK}
H. Arenh\"{o}vel and M. Sanzone, \textit{Photodisintegration of the Deuteron: a Review of Theory and Experiment} (Springer-Verlag, Berlin, 1991).
\bibitem{chen99}
J. W. Chen and M. J. Savage, \href{https://doi.org/10.1103/PhysRevC.60.065205}{Phys. Rev. C {\bf 60}, 065205 (1999)}.
\bibitem{joh01}
A. S. Johnson and G. M. Hale, \href{https://doi.org/10.1016/S0375-9474(01)00789-8}{Nucl. Phys. A {\bf 688}, 566 (2001)}.
\bibitem{rup00}
G. Rupak, \href{https://doi.org/10.1016/S0375-9474(00)00323-7}{Nucl. Phys. A {\bf 678}, 405 (2000)}.
\bibitem{wang22} 
H. W. Wang  \textit{et al.}, \href{https://doi.org/10.1007/s41365-022-01076-0}{Nucl. Sci. Tech. {\bf 33}, 87 (2022)}.
\bibitem{xu22} 
H. Xu \textit{et al.}, \href{https://doi.org/10.1016/j.nima.2022.166742}{Nucl. Instrum. Meth. A {\bf 1033}, 166742 (2022)}.
\bibitem{liu24} 
L. X. Liu \textit{et al.}, \href{https://doi.org/10.1007/s41365-024-01469-3}{Nucl. Sci. Tech. {\bf 35}, 111 (2024)}.
\bibitem{he14} 
J. He \textit{et al.}, \href{https://doi.org/10.1093/nsr/nwt039}{Natl. Sci. Rev. {\bf 1}, 171 (2014)}.
\bibitem{xu2025}
H. H. Xu \textit{et al.}, \href{https://doi.org/10.1016/j.nima.2025.170249}{Nucl. Instrum. Meth. A {\bf 1073}, 170249 (2025)}.
\bibitem{hao21} 
Z. Hao \textit{et al.}, \href{https://doi.org/10.1016/j.nima.2021.165638}{Nucl. Instrum. Meth. A {\bf 1013}, 165638 (2021)}.
\bibitem{hao25} 
Z. R. Hao \textit{et al.}, \href{https://doi.org/10.1016/j.scib.2025.05.037}{Sci. Bull. {\bf 70}, 2591 (2025)}.
\bibitem{supp}
See Supplemental Material at http://link.aps.org/supplemental/10.1103/tbbt-s819 for more details on the $\gamma$-ray beam spectra, photoneutron cross sections, and thermonuclear reaction rates which includes Refs. [9,13,29,33].
\bibitem{hao20} 
Z. Hao \textit{et al.}, \href{https://dx.doi.org/10.11889/j.0253-3219.2020.hjs.43.110501}{Nucl. Tech. (in Chinese), {\bf 43}, 110501 (2020)}.
\bibitem{ren18} 
T. Renstrøm \textit{et al.}, \href{https://doi.org/10.1103/PhysRevC.98.054310}{Phys. Rev. C {\bf 98}, 054310 (2018)}.
\bibitem{ame20} 
M. Wang \textit{et al.}, \href{https://iopscience.iop.org/article/10.1088/1674-1137/abddaf/meta}{Chin. Phys. C {\bf 45}, 030003 (2021)}.
\bibitem{hao25_1}
Z. R. Hao \textit{et al.}, \href{https://doi.org/10.1007/s41365-025-01773-6}{Nucl. Sci. Tech. \textbf{36}, 183 (2025)}.
\bibitem{haoPRL_sub}
Z. R. Hao \textit{et al.}, to be submitted.
\bibitem{nn-online}
NN-OnLine, \href{https://nn-online.org/}{https://nn-online.org/}.
\bibitem{DA13}
B. H. Daub \textit{et al.}, \href{https://doi.org/10.1103/PhysRevC.87.014005}{Phys. Rev. C {\bf 87}, 014005 (2013)}.
\bibitem{uts15} 
H. Utsunomiya \textit{et al.}, \href{https://doi.org/10.1103/PhysRevC.92.064323}{Phys. Rev. C {\bf 92}, 064323 (2015)}.
\bibitem{Kawano1992}
L. Kawano, \href{https://ntrs.nasa.gov/api/citations/19920015920/downloads/19920015920.pdf}{NASA STI/Recon Technical Report {\bf N92}, 25163 (1992)}.
\bibitem{Smith:1992yy}
M. Smith, L. Kawano, and R. Malaney, \href{https://adsabs.harvard.edu/full/1993ApJS...85..219S}{Astrophys. J. Suppl. Ser., {\bf 85}, 219 (1993)}.
\bibitem{ParticleDataGroup:2018ovx}
M. Tanabashi \textit{et al.} (Particle Data Group), \href{https://doi.org/10.1103/PhysRevD.98.030001}{Phys. Rev. D {\bf 98}, 030001 (2018)}.
\bibitem{Pisanti:2020efz}
O.~Pisanti, G.~Mangano, G.~Miele, and P.~Mazzella, \href{https://iopscience.iop.org/article/10.1088/1475-7516/2021/04/020/meta}{J. Cosmol. Astropart. Phys. {\bf 04}, 020 (2021)}.
\bibitem{man05}
G. Mangano {\it et al.}, \href{https://www.sciencedirect.com/science/article/pii/S0550321305008291}{Nucl. Phys. B {\bf 729}, 221 (2005)}.
\bibitem{sal16}
P. De Salas and S. Pastor, \href{https://dx.doi.org/10.1088/1475-7516/2016/07/051}{J. Cosmol. Astropart. Phys. {\bf 07}, 051 (2016)}.
\bibitem{LUNA20}
V. Mossa \textit{et al.}, \href{https://doi.org/10.1038/s41586-020-2878-4}{Nature {\bf 587}, 210 (2020)}.
\bibitem{Pisanti:2007hk}
O. Pisanti \textit{et al.}, \href{https://www.sciencedirect.com/science/article/pii/S0010465508000921}{Comput. Phys. Commun. {\bf 178}, 956 (2008)}.
\bibitem{coc15}
A. Coc \textit{et al.}, \href{https://journals.aps.org/prd/pdf/10.1103/PhysRevD.92.123526}{Phys. Rev. D {\bf 92}, 123526 (2015)}.
\bibitem{gom17}
A. G\'omez I\~nesta \textit{et al.}, \href{https://dx.doi.org/10.3847/1538-4357/aa9025}{\apj, {\bf 849}, 134 (2017)}.
\bibitem{Planck:2018vyg}
N.~Aghanim \textit{et al.} (Planck Collaboration), \href{https://doi.org/10.1051/0004-6361/201833910}{Astron. Astrophys. {\bf 641}, A6 (2020).}



\end{thebibliography}
\end{document}